# Effect of externally applied resonant magnetic perturbations on resistive tearing modes

**Qiming Hu[1], Q. Yu[2], Bo Rao[1], Yonghua Ding[1], Xiwei Hu[1], Ge Zhuang[1,*],** and the J-TEXT Team1

[1] State Key Laboratory of Advanced Electromagnetic Engineering and Technology, Huazhong University of Science and Technology, Wuhan, 430074, China

[2] Max-Planck-Institut für Plasmaphysik, EURATOM association, 85748 Garching, Germany

* Correspongding author: ge-zhuang@mail.hust.edu.cn

**Abstract.** Static resonant magnetic perturbations (RMPs) generated by saddle coil current have been applied in J-TEXT tokamak experiments in order to study their effects on tearing mode instabilities. With increasing the RMP amplitude in time during the discharge, the mode stabilization is first observed, but a large locked mode follows if the RMP amplitude is increased to a too large value, indicating that the RMP amplitude is important in determining the plasma response and the tearing mode behavior. By careful adjustment of the RMP amplitude, the (partial) stabilization of the m/n =2/1 tearing mode by RMPs of moderate amplitude has been achieved without causing mode locking ($m$ and $n$ are the poloidal and toroidal mode numbers). To compare with experimental results, nonlinear numerical modeling based on reduced MHD equations has been carried out. With experimental parameters as input, both the mode locking and mode stabilization by RMPs are also obtained from numerical modeling. Further calculations have been carried out to study the plasma parameters affecting the mode stabilization by RMPs, including the plasma rotation frequency, viscosity, Alfvén velocity, and the RMPs amplitude. It is found that the suppression of the tearing mode by RMPs of moderate amplitude is possible for a sufficiently high ratio of plasma rotation velocity to the Alfvén speed. A larger plasma viscosity enhances the mode stabilization.

## 1.Introduction

It is well known that tearing instabilities, driven by unfavorable plasma current density gradient, generate magnetic islands at their corresponding rational surfaces in tokamak plasmas [1-7]. The low-$m$ tearing modes, such as the $m/n$=3/2 and 2/1 modes ($m$ and $n$ are the poloidal and toroidal mode numbers, respectively), often lead to a degradation of plasma confinement. The 2/1 mode can even stop rotation due to the locking effect of the vacuum vessel and error field and cause major disruptions [8-10]. The stabilization of (neoclassical) tearing modes is therefore an important issue for a fusion reactor.

The width of a sufficiently large magnetic island is known to be determined by plasma current density and safety factor profiles and the bootstrap current perturbation proportional in part to plasma beta value. When the island width is not too large, however, a variety of stabilizing or destabilizing mechanisms are involved, such as the ion polarization current [11-16] and the plasma viscosity [14, 17]. The classical tearing mode stability in low β plasmas has been investigated in [18]. In recent years, the effect of plasma flow and its shear on the tearing mode stability were also studied [19-22]. As complicated mechanisms are involved in determining the mode amplitude, a better understanding of the island physics is necessary.

On the other hand, the intrinsic error field of tokamak or externally applied resonant magnetic perturbations (RMPs) can significantly affect the tearing modes [17, 23-42]. For the plasma being originally stable to tearing modes, an applied RMP can penetrate through to the rational surface and generate a magnetic island there [17, 24, 35, 38]. Once an island is sufficiently large, it will be locked to the machine error field or applied RMPs [17, 23-25]. In addition, RMPs of moderate amplitude were found to reduce the island size in tokamak experiments [17, 24, 39-42]. These different findings have attracted great interest in fusion research, and further investigation is required in order to understand the plasma response to RMPs, especially when considering of the stabilization of edge localized modes (ELMs) by RMPs in recent years and its future application in ITER [43-45].

Recently static RMPs, generated by saddle coil current, were utilized to study their effect on MHD instabilities in J-TEXT tokamak [46] (former TEXT-U). The experimental results reveal that in addition to the mode locking caused by sufficiently large RMPs, the $m/n$=2/1 tearing modes can be (partly) stabilized by RMPs of a moderate amplitude. Motivated by experimental results, the effect of RMPs on the $m/n$=2/1 tearing mode is studied numerically by using the nonlinear (reduced) MHD equations. With experimental parameters as input, the mode stabilization is also seen from numerical results. Detailed calculations have been further carried out to understand the plasma parameters affecting the mode stabilization, including the plasma rotation frequency, viscosity and the Alfvén velocity. It is found that the stabilization of the 2/1 mode by RMPs is possible for a sufficiently fast rotating plasma and low Alfvén velocity.

Our experimental setup and results are introduced in section 2. In section 3 the theoretical model and numerical results are presented, and the discussion and summary are given in section 4.

## 2. Experimental results

On J-TEXT tokamak there are a set of saddle coils toroidally distributed to generate static RMPs, similar to those used on TEXT-U [47], leading to a resonant $m/n$=2/1 component of radial magnetic field $b_r$ of strength 63$\mu$T at plasma edge $r=a$ per kilo-ampere of coil current. The location of the saddle coils, consisting of 5 coils at different toroidal angles labeled as sine1-2 and cosine1-3, are shown in Fig. 1a. The spectrum of the generated $b_r$ of the resonant hand at plasma edge $r=a$, corresponding to the studies presented in this paper, are shown in Fig.1b for one kilo-ampere of the saddle coil current, from which it is seen that resonant $m/n$=2/1 and 3/1 components have larger amplitude for $q_a$<3.5, where $q_a$ is the edge safety factor at r=a. Other components, such as the $m/n$=3/2 component, have much smaller amplitude.

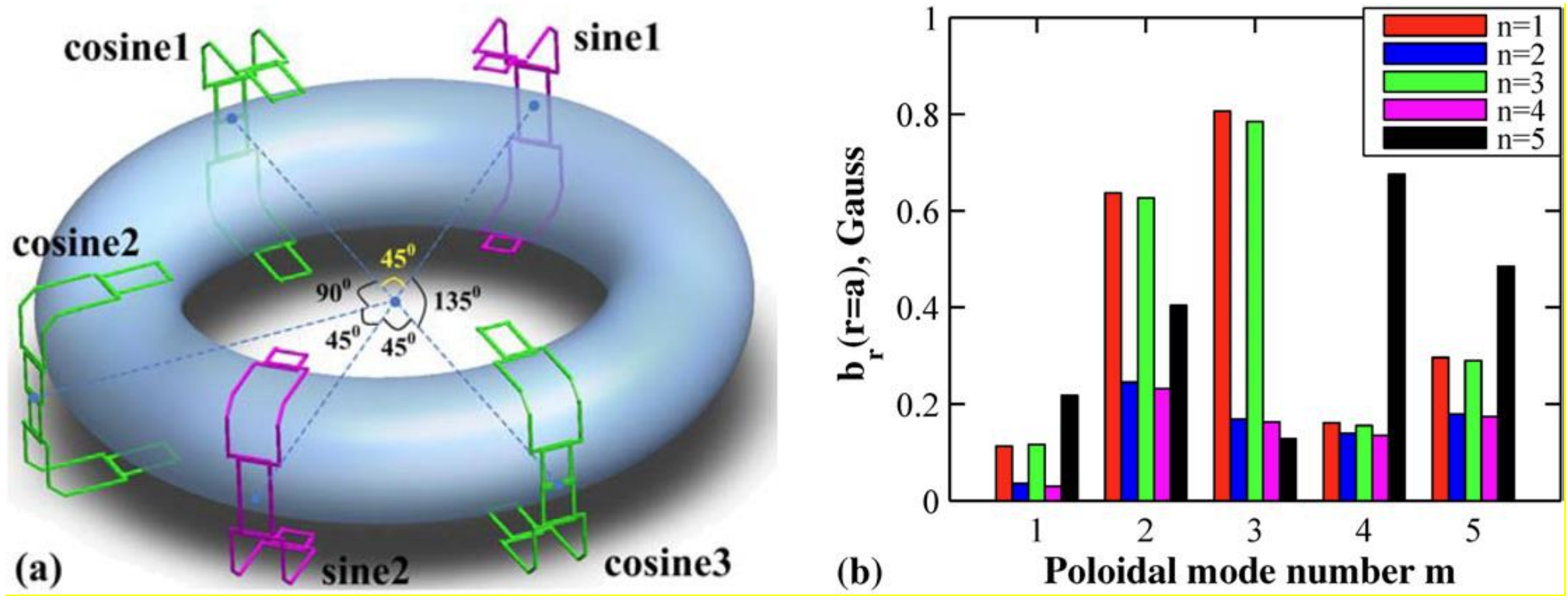

**Figure 1.** (a) Layout of saddle coils. (b) Spectrum of the generated radial magnetic field $b_r$ of the resonant hand at plasma edge at $r=a$=0.27m.

On J-TEXT many experimental shots have been devoted to study the effect of RMPs on the tearing mode instability. Two different types of plasma response have been observed: (a) when a sufficiently large RMP (i.e. $I_{coil}$>5.5kA or $b_r$($m/n$=2/1)>3.5 Gauss) is applied during the discharges, it leads to the locking of an existing rotating $m/n$=2/1 tearing mode as expected, resulting in a larger magnetic island and confinement degradation; (b) when a RMP of moderate amplitude (i.e. 3.5kA<$I_{coil}$<5.5kA or 2.2 Gauss<$b_r$($m/n$=2/1)<3.5 Gauss) is applied, it can, however, (partly) stabilize the $m/n$=2/1 tearing mode. In order to evaluate which Fourier component affects the $m/n$=2/1 tearing mode, experiments have also been carried out with other RMPs spectrum, obtained by using different combination of the current phase and amplitude for each individual coil. For the case that the amplitude of the resonant $m/m$=3/1 or other component is dominant and much larger than the 2/1 component, no obvious effect of RMPs on the 2/1 tearing mode has been observed. Only when the 2/1 component has a large amplitude as shown in figure 1(b), the effect of RMPs on the 2/1 mode is clearly seen.

Two typical experimental examples are shown in the following figures 2 and 3. In these Ohmic discharges the toroidal magnetic field is $B_t$=1.8T at magnetic axis, the plasma current $I_p$=180kA, the line average electron density at the central channel is $n_e=1.5\times10^{19}$m$^{-3}$, and the plasma minor and major radius are $a$=0.27m and $R$=1.05m, respectively. The safety factor $q$ at plasma edge is 3.5, the volume averaged $\beta_p=\langle P_e\rangle/(B_\theta^2/2\mu_0)$ at the plasma surface is 0.1, the original electron diamagnetic drift frequency $f_{*e}$ for the 2/1 mode is about 1.5kHz, and the $m/n$=2/1 tearing mode propagates in the electron diamagnetic drift direction.

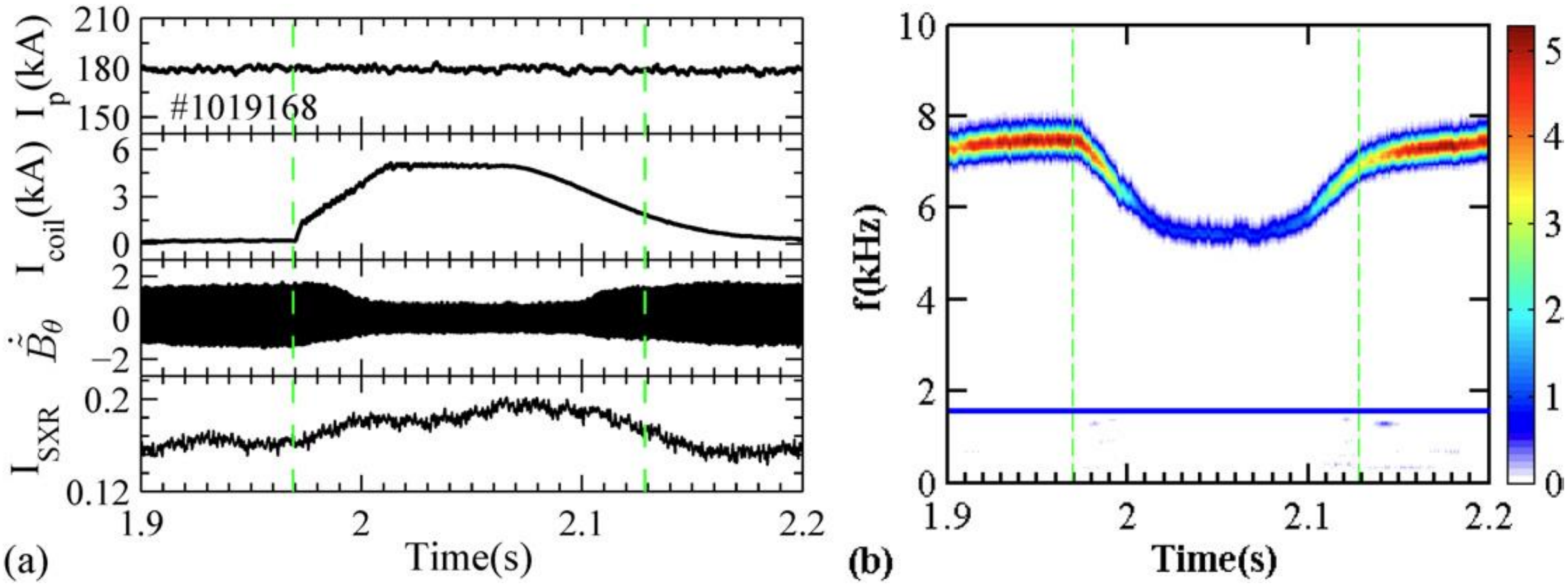

**Figure 2.** (a) Time evolution of plasma parameters during discharge #1019168. From the top, plasma current $I_p$, RMPs coils current $I_{coil}$, Mirnov signal d$B_\theta$/d$t$, and intensity of soft X-ray emission $I_{sxr}$. (b) Corresponding wavelet power spectrum of Mirnov signal to show the mode frequency and strength. The horizontal solid line indicates the original equilibrium electron diamagnetic drift frequency of the 2/1 mode, $f_{*e}$=1.5kHz. The mode is partly stabilized by RMPs in the time period between two vertical dashed lines. $I_{coil}$=4.95kA in the current flattop. When $I_{coil}$ is reduced to 3.5kA at $t$=2.1s, the mode amplitude and frequency increase again.

The time evolution of measured parameters (shot #1019168), the plasma current $I_p$, RMPs coils current $I_{coil}$, Mirnov signal d$B_\theta$/d$t$, and intensity of soft X-ray emission (SXR) $I_{sxr}$, are presented in figure 2(a). In figure 2(b) the corresponding wavelet power spectrum (a.u.) of the Mirnov signal is given, showing the time evolution of the mode frequency and strength. The horizontal solid line indicates the original equilibrium electron diamagnetic drift frequency of the 2/1 mode, $f_{*e}$=1.5kHz. When $I_{coil}$ is increased to 4.95kA in the current flattop, the Mirnov signal amplitude |d$B_\theta$/d$t$| is decreased by more than 50%, and the mode frequency is decreased from 7.5 kHz to 5.5 kHz. Thus the decrease of more than 50% in |d$B_\theta$/d$t$| is not accountable by the decrease in mode frequency of 27%. Meanwhile $I_{sxr}$ increases, indicating the improved confinement due to the decreased island width. In J-TEXT, 8 arrays are installed in order to measure the intensity of SXR emission, and each array contains 16 viewing chords covering the plasma cross-section. The SXR emission shown in figure 2 and figure 3 is measured by the viewing chord with chord radius $r$=20cm, being about the radial location of the $q$=2 surface. Also, the intensity of SXR emission inside the $q$=2 surface shows a similar increase during mode stabilization by RMPs. The data analysis reveals that the dominant mode is the $m/n$=2/1 mode in the discharge, and it rotates in the electron diamagnetic drift direction. When $I_{coil}$ is decreased to 3.5kA at 2.1s, the mode amplitude and frequency increase again. With further decreasing $I_{coil}$ , the mode amplitude and frequency recover to their previous values before applying RMPs. The mode is partly stabilized by RMPs in the time period between two vertical dashed lines.

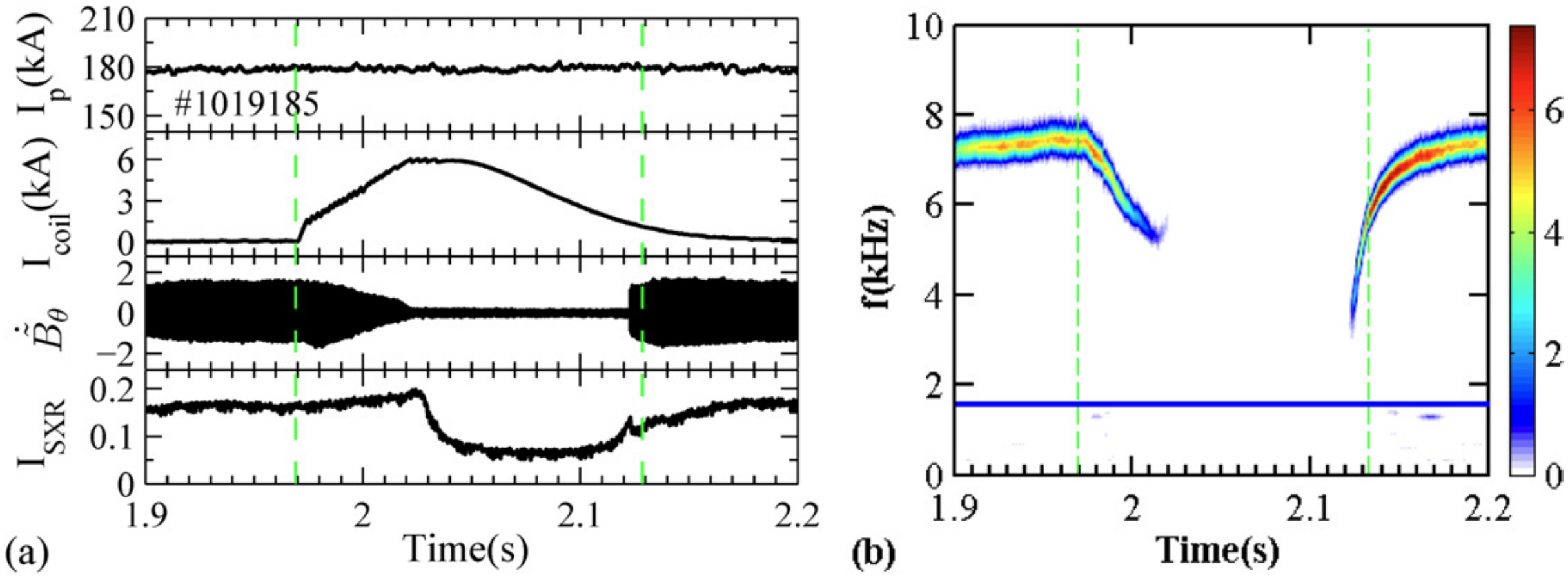

**Figure 3.** Similar to figure 2 but with larger RMPs amplitude. It is shown (a) the time evolution of plasma parameters during discharge #1019185 and (b) corresponding wavelet power spectrum of Mirnov signal. The 2/1 mode is first suppressed by RMPs, but mode locking follows shortly later, resulting in a large island and a corresponding significant reduction in the intensity of SXR emission $I_{sxr}$. The mode unlocks from RMPs when the coil current is ramped down to a sufficiently small value. The horizontal line indicates the original equilibrium electron diamagnetic drift frequency of the 2/1 mode with $f_{*e}$=1.5kHz. The mode amplitude and frequency are affected by RMPs in the time period between the two vertical dashed lines.

Figure 3 (discharge 1019185) shows another example with larger amplitude of applied RMPs. When the RMPs coil current is increased from zero to the flattop $I_{coil}$=6kA at 2.02s, the tearing mode is first stabilized to a very small amplitude. The mode amplitude decreases faster than frequency decreases. The intensity of SXR emission increases in this period of time. However, mode locking happens 6ms later, resulting in a large locked island, as indicated by the significant reduction of $I_{sxr}$ and the time trace of the mode frequency shown in 3(b). When the RMPs coil current is reduced to 1.4kA at 2.12s, the mode unlocks from the applied RMPs. The observed mode frequency begins from about 3.75 kHz in the unlocking phase, being about half of the mode frequency before applying RMPs. The required RMPs amplitude for mode unlocking is smaller than that for causing mode locking as expected [17, 24]. With further decreasing the RMPs amplitude, the mode amplitude and frequency recover to their original values before applying RMPs. In the time period between the two vertical dashed lines, the mode amplitude and frequency are affected by RMPs.

## 3. Numerical modeling

In order to understand and compare with the experimental results as shown in figures 2 and 3, numerical modeling has been carried out using the low-β and large tokamak aspect ratio approximation. The magnetic field is defined as

, (1)

where $\psi$ is the helical flux function, $m/r$ and $k=n/R$ are the wave vectors in the $\boldsymbol{e}_\theta$ (poloidal ) and $\boldsymbol{e}_t$ (toroidal) directions, and r and $R$ are the minor and the major radius, respectively.

Ohm's law and the equation of motion (after taking the operator $\boldsymbol{e}_t \cdot \nabla \times$),

, (2)

, (3)

are utilized, where

, (4)

$\boldsymbol{v}$ is the plasma velocity,

(5)

is the toroidal plasma current density, $\eta$ is the plasma resistivity, $\rho$ the mass density, and $\mu$ the plasma viscosity. The stream function $\phi$ is defined by

, (6)

and the subscripts || and ⊥ stand for the parallel and the perpendicular components, respectively. $E_0$ is the equilibrium electric field for maintaining the original equilibrium plasma current density, and $S_m$ in equation (3) is the momentum source, which leads to a poloidal equilibrium plasma rotation.

Normalizing the length to the plasma minor radius $a$, the time $t$ to $\tau_R$, $\psi$ to $aB_t$ , $\phi$ to $a^2/\tau_R$, velocities to $a/\tau_R$, and $j$ to $B_t/a$ , Eqs. (2) and (3) become

, (7)

, (8)

where $\tau_R= a^2\mu_0/\eta$ is the resistive time, $S=\tau_R/\tau_A$ the magnetic Reynolds number, $\tau_A=a/V_A$ the Alfvén time, $V_A=B_t/(\mu_0\rho)^{1/2}$ the Alfvén velocity, and $U=-\nabla_\perp^2\phi$ the plasma vorticity.

The effect of a single helicity RMP with $m/n$=2/1 is taken into account by the boundary condition

, (9)

where $\psi_a$ describes the normalized $m/n$=2/1 helical magnetic flux amplitude at $r=a$, and $\theta$ and $\varphi$ are the poloidal and toroidal angle, respectively. The radial magnetic field perturbation at $r=a$ is given by $b_{ra}=m\psi_a B_t$.

Equations (7) and (8) are solved simultaneously using the initial value code TM1, which has been used for modeling the nonlinear growth and saturation of NTMs and their stabilization by RF current earlier [36-37]. It should be mentioned that the diamagnetic drift has not been included in our model, which is important in determining the tearing mode stability for high $\beta$ plasmas with large electron pressure gradient [11-16, 48-50]. Linear tearing modes are found to be completely stabilized for a sufficiently large electron diamagnetic drift frequency [11, 48-49]. In the nonlinear phase the ion polarization current associated with diamagnetic drifts leads to a threshold for the onset of neoclassical tearing modes [11-16]. For J-TEXT Ohmic plasmas, however, the $\beta$ value is low. The electron diamagnetic frequency for the 2/1 mode is about $f_{*e}\cong1.5$kHz, being significantly lower than the mode frequency $f$~7.5kHz. Also, the effect of diamagnetic drifts is weak due to the pressure flattening across a large island [11].

**3.1. Comparison with experimental results**

To compare with experimental results shown in Section 2, the input parameters for following calculations are based on J-TEXT experimental parameters. A monotonic profile for the safety factor $q$ is taken with the $q$=2/1 surface located at $r_s$=0.7$a$. The electron temperature is about 300eV at the $q$=2 surface, and the local electron density is about $1\times10^{19}$m$^{-3}$. Plasma viscosity is assumed to be at the anomalous transport level $\mu_\perp$=0.5m$^2$s$^{-1}$. These parameters lead to S=$6\times10^6$ and $\mu_\perp$=0.82($a^2/\tau_R$).

The mode rotation observed in J-TEXT experiments is along the toroidal direction. In our theoretical model, however, only the poloidal plasma rotation is included, so that a larger plasma viscosity, $\mu_\perp$=70($a^2/\tau_R$), is utilized in the following calculations for a reasonable balance between the electromagnetic and viscous force, because (a) the electromagnetic force to slow down the island rotation in the toroidal direction is smaller by a factor $(n/m)(r_s/R)$ compared with that in the poloidal direction; (b) to obtain the same mode frequency due to the plasma rotation, the toroidal rotation speed should be $(m/n)(R/r_s)$ times larger than the poloidal one. These two effects lead to a relative larger viscous force for the toroidal rotation case by a factor of $[(m/n)(R/r_s)]^2$ [24], which is of the order of $10^2$.

In figure 4 of the calculation results, the time evolution of the normalized mode angular frequency, $\omega_{1/2}\tau_R$ (a), and island width, $w/a$ (b), are shown. The plasma is unstable to the 2/1 tearing mode for $\psi_a$=0, so the magnetic island grows and saturates at a width of $w$=0.07a before the RMP is applied. The original mode frequency before applying RMP is $\omega_0=5.5\times10^3/\tau_R$ (7.5kHz). The RMP waveform utilized in the calculation is also shown in (b). In the flattop $\psi_a=8.5\times10^{-5}aB_t$, corresponding to $b_r(r=a)=m\psi_aB_t$=3.06 Gauss ($I_{coil}$=4.9kA) of figure 2. For $1.075\tau_R\leq t\leq1.4\tau_R$, the RMP amplitude linearly increases with time, leading to the decrease in both the mode frequency and island width. The RMP amplitude reaches the flattop at $t=1.4\tau_R$ and is kept constant until $t=2.235\tau_R$. During this period of time, the island width is decreased to 0.05$a$, and the mode frequency oscillates around a constant value $\omega=4\times10^3/\tau_R$, being smaller than the original mode angular frequency. The detailed time evolution of the mode frequency during this phase is shown at the right corner of figure 4(a). For $2.235\tau_R\leq t\leq2.8\tau_R$, the RMP amplitude decreases in time, resulting in an increase in both the mode frequency and island width. The RMP is turned off at $t$=2.8$\tau_R$, so that the mode frequency and amplitude recover to their original values. The decrease in both the mode amplitude and frequency by RMP is seen from figure 4, being essentially in agreement with the corresponding experimental observation shown in figure 2.

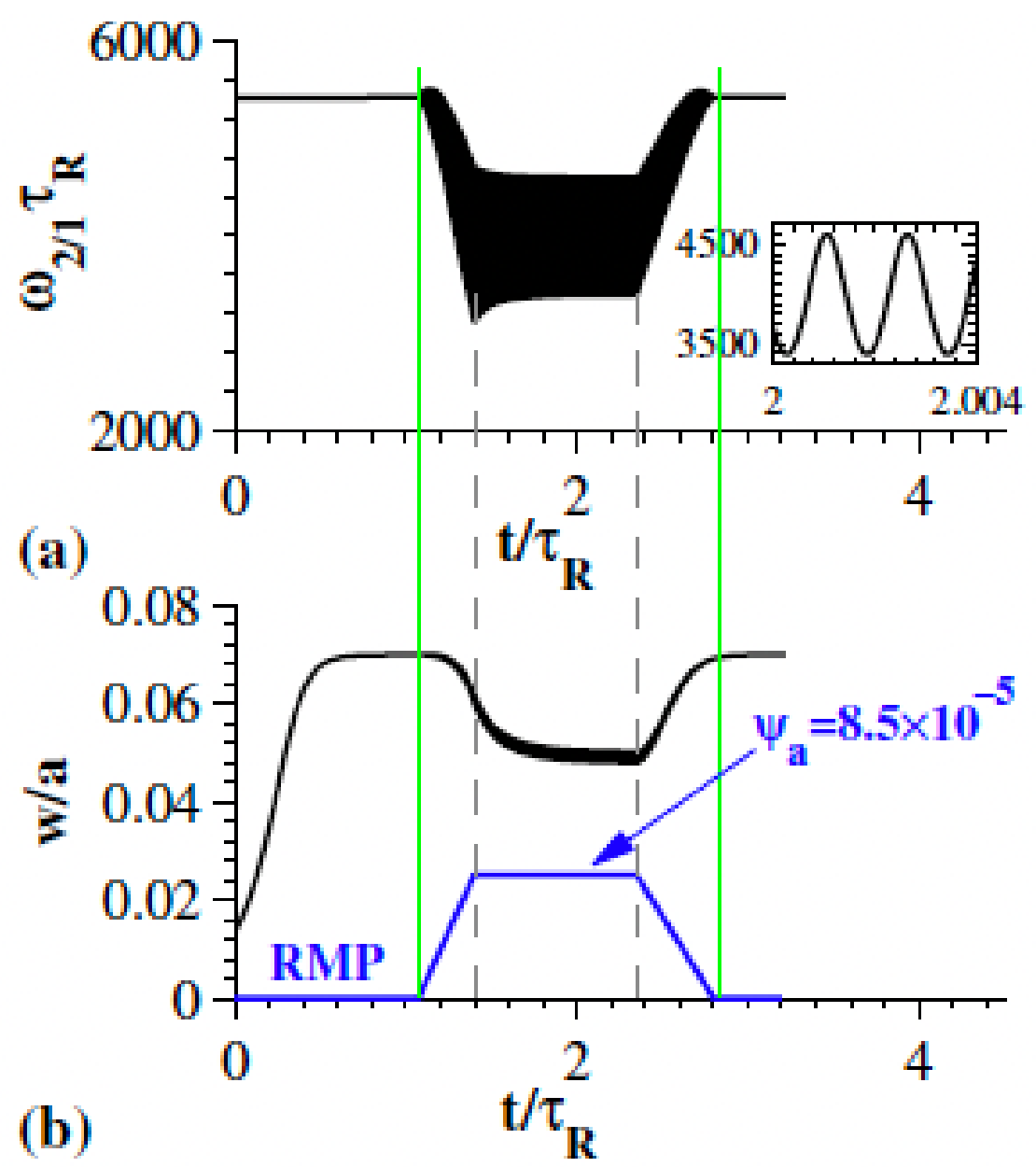

**Figure 4.** Time evolution of the normalized mode angular frequency $\omega_{1/2}\tau_R$ (a) and island width $w/a$ (b). The applied RMP waveform is also shown in (b). The 2/1 tearing mode is partly stabilized by applied RMP as seen from the decreased island width, and the mode frequency decreases during the mode stabilization period. The numerical results approximately agree with the corresponding experimental observation shown in figure 2.

**Figure 5.** Similar to figure 4 but with a larger RMP amplitude, the time evolution of the normalized mode angular frequency $\omega_{1/2}\tau_R$ (a) and island width $w/a$ (b). The applied RMP waveform is also shown in (b). The 2/1 mode is first suppressed by but then locked to RMP, resulting in a large locked island. The numerical results approximately agree with the corresponding experimental observation shown in figure 3.

For larger RMP amplitude, the time evolution of the normalized angular mode frequency $\omega_{1/2}\tau_R$ (a) and island width $w/a$ (b) are shown in figure 5. The applied RMP waveform is also shown in (b). In the flattop $\psi_a$=1.1×10$^{-4}aB_t$, corresponding to $b_r(r=a)$=3.96 Gauss ($I_{coil}$=6.2kA) of figure 3. The applied RMP amplitude first linearly increases in time, resulting in the decrease in both the mode frequency and island width, similar to those seen from figure 4. The mode frequency decreases by about 50% of $\omega_0$ at $t$=1.4$\tau_R$, and then the island width quickly increases to a large value $w$=0.172a in a very short time (about 0.01$\tau_R$). Meanwhile, the mode frequency drops to zero, being typical for mode locking [27]. The detailed evolution of island width just before and after mode locking is shown at the right corner of figure 5(b). For 1.8$\tau_R \leq t \leq$2.3$\tau_R$, RMP decreases in time, leading to the decrease in the island width, but the mode remains to be locked until $t$=2.22$\tau_R$. After mode unlocking, the island width quickly decreases. The time evolution of mode frequency just before and after mode unlocking is shown at the right corner of figure 5(a). The RMP is turned off at $t$=2.3$\tau_R$, so that afterwards the mode frequency and island width gradually recover to their original values. The numerical results approximately agree with the corresponding experimental results shown in figure 3.

### 3.2. Effect of plasma parameters

In order to have a general understanding about the effect of RMPs on the tearing mode, extensive numerical modeling has been further carried out, to study the effect of plasma parameters affecting the mode stabilization or mode locking. For the following results $S=1\times10^7$ and a small plasma viscosity, $\mu_\perp=2.8(a^2/\tau_R)$, is taken. Other parameters are kept the same as those in Section 3.1 except mentioned elsewhere.

To look into the effect of plasma rotation frequency, in figure 6 the normalized mode angular frequency, $\omega_p/\omega_0$, and island width at nonlinear saturation are shown as a function of the applied RMP amplitude $\psi_a$ for $\omega_0=6.3\times10^3/\tau_R$ in figure 6(a) and $9.5\times10^3/\tau_R$ in figure 6(b), respectively, where $\omega_p$ ($\omega_0$) is the mode angular frequency with (without) the RMP. The initial island width for these calculations is $w_0=0.04a$ at $t=0$, and the RMP is applied from $t=0$. For a lower original mode frequency ($\omega_0=6.3\times10^3/\tau_R$ in figure 6(a)), no mode stabilization by the RMP is observed. The mode frequency decreases with increasing $\psi_a$, while the island width slightly increases. The mode locking happens for sufficiently large RMP amplitude, $\psi_a>4.1\times10^{-6}aB_t$, resulting in a sharp increase in the island width and drop in the mode frequency (to zero). The about 50% drop in mode frequency before mode locking has been predicted theoretically in [27] and observed experimentally in [24]. For a higher original mode frequency ($\omega_0=9.5\times10^3/\tau_R$ in figure 6(b)), however, the mode is stabilized by the RMP if the RMP amplitude is not too large. Three different regimes are observed from figure 6(b):

(i) *Mode suppression regime* ($\psi_a<4.46\times10^{-5}aB_t$), in which the island width decreases with increasing RMP amplitude. It is interesting to note that the mode frequency decreases with increasing $\psi_a$ for $\psi_a\leq2.2\times10^{-5}aB_t$ but increases for $2.2\times10^{-5}aB_t<\psi_a<4.46\times10^{-5}aB_t$. Such a frequency increase is likely to be caused by a weaker electromagnetic force acting on the island with decreasing island width.

(ii) *Small locked island regime* ($4.46\times10^{-5}aB_t<\psi_a<5.6\times10^{-5}aB_t$), in which the mode is locked to the RMP as indicated by the zero mode frequency, while the island width is very small, being different from the usual mode locking which corresponds to a large island;

(iii) *mode locking regime* ($\psi_a>5.6\times10^{-5}aB_t$), beginning from a large jump in the island width. The mode frequency is also zero in this regime, but the island width is significantly larger than the original one without RMP.

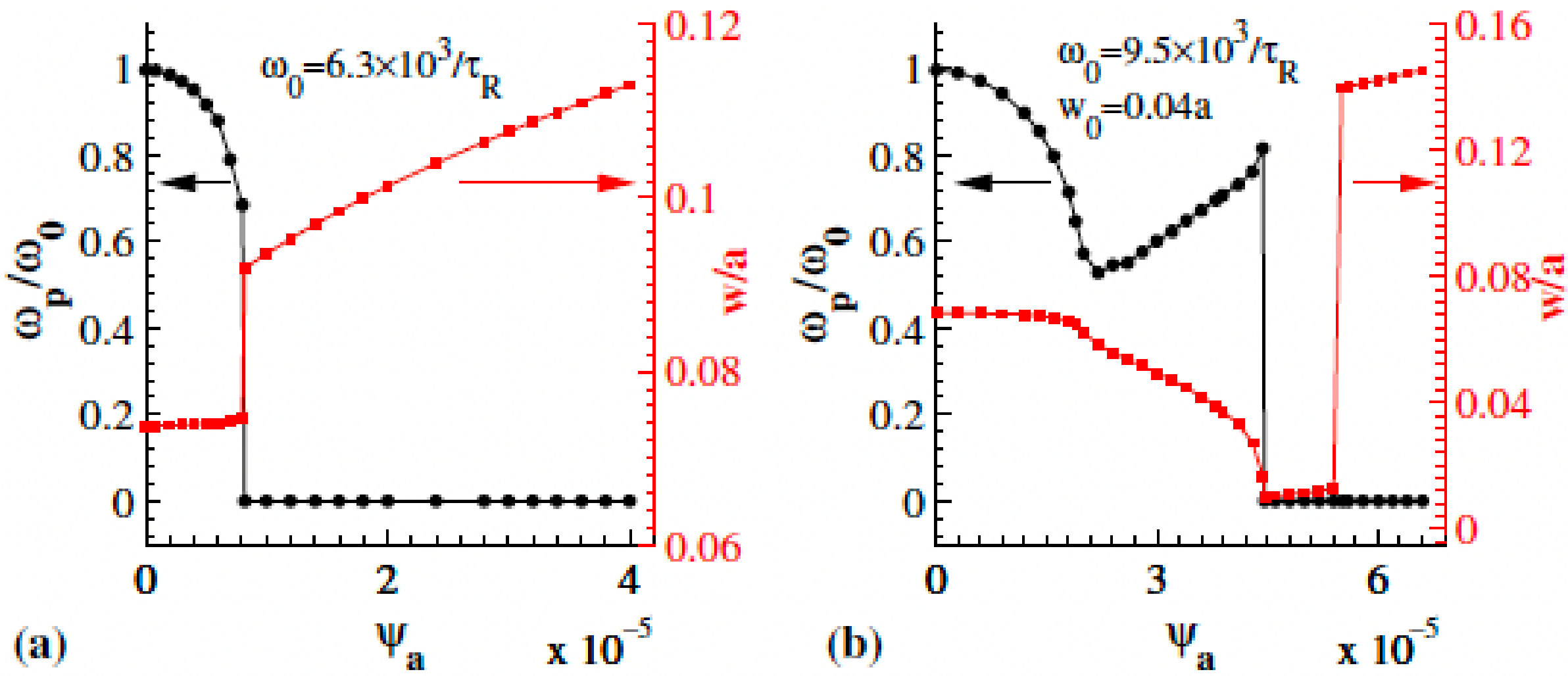

**Figure 6.** Normalized mode angular frequency, $\omega_p/\omega_0$, and island width, $w/a$, at nonlinear saturation versus $\psi_a$ for $\omega_0=6.3\times10^3/\tau_R$ (a) and $9.5\times10^3/\tau_R$ (b), with the initial island width $w_0=0.04a$. For a lower value of $\omega_0$ as shown in (a), only mode locking is seen with increasing $\psi_a$. For a higher value of $\omega_0$ (b), however, three different regimes exist: (i) *Mode suppression regime* ($\psi_a<4.46\times10^{-5}aB_t$), in which the island width decreases with increasing RMP amplitude; (ii) *Small locked island regime* ($4.46\times10^{-5}aB_t<\psi_a<5.6\times10^{-5}aB_t$), in which the mode is locked to the RMP as indicated by the zero mode frequency, while the island width is quite small; (iii) *Mode locking regime* ($\psi_a>5.6\times10^{-5}aB_t$), beginning from a large jump in the island width.

Corresponding to figure 6(b), the radial profiles of the $m/n$=0/0 component of the poloidal plasma velocity, $v_\theta(a/\tau_R)$, at nonlinear saturation are shown in figure 7 for (a) $\psi_a=4\times10^{-5}aB_t$ and (b) $\psi_a=5\times10^{-5}aB_t$ and $\psi_a=6\times10^{-5}aB_t$. The dashed curve shows the original radial profile of $v_\theta$ at $t$=0, and the vertical line at $r=0.7a$ indicates the radial location of the $q$=2 rational surface. For $\psi_a=4\times10^{-5}aB_t$ in the *mode suppression regime*, the local poloidal plasma velocity around the rational surface oscillates in time, and no steady $v_\theta$ profile is found. The flow shear around the island significantly changes. For $\psi_a=5\times10^{-5}aB_t$ in the *small locked island regime* shown in figure 7(b), the plasma velocity decreases from its original value but remains finite at the rational surface, since the island width is sufficiently small in this case to allow the plasma to slip through it. For $\psi_a=6\times10^{-5}aB_t$ in the *mode locking regime*, the plasma velocity is brought to zero at the rational surface as expected, being different from that in the small locked island regime.

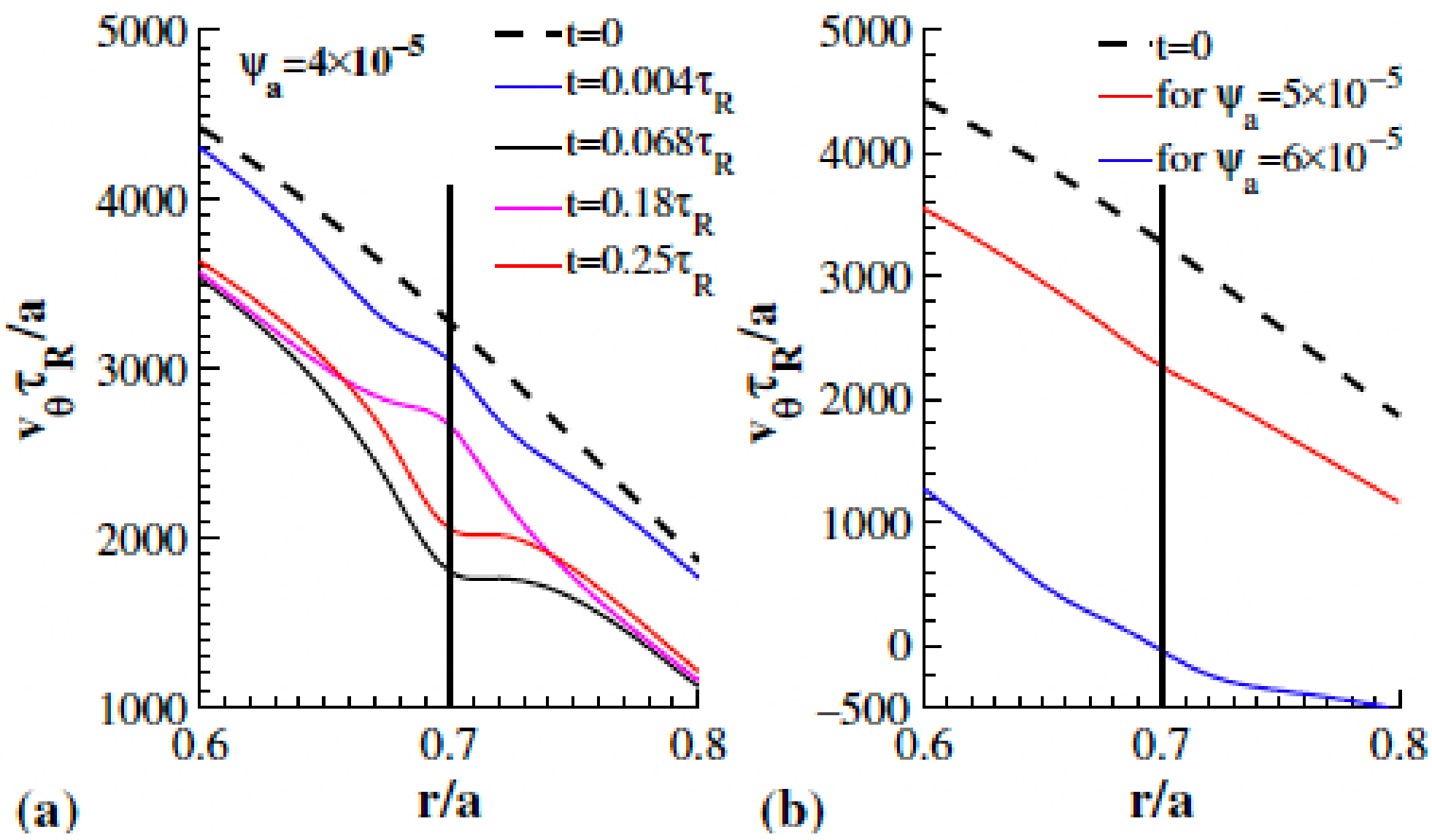

**Figure 7.** Corresponding to figure 6(b), radial profiles of the $m/n$=0/0 component of the normalized poloidal plasma rotation velocity, $v_\theta(a/\tau_R)$, at nonlinear saturation for (a) $\psi_a$=4×10$^{-5}aB_t$ in the *mode suppression regime* and (b) $\psi_a$=5×10$^{-5}aB_t$ in the *small locked island regime* and $\psi_a$=6×10$^{-5}aB_t$ in the *mode locking regime*. The dashed curve shows the original radial profile of $v_\theta$ at $t$=0, and the vertical line at $r$=0.7$a$ shows the radial location of the $q$=2 rational surface.

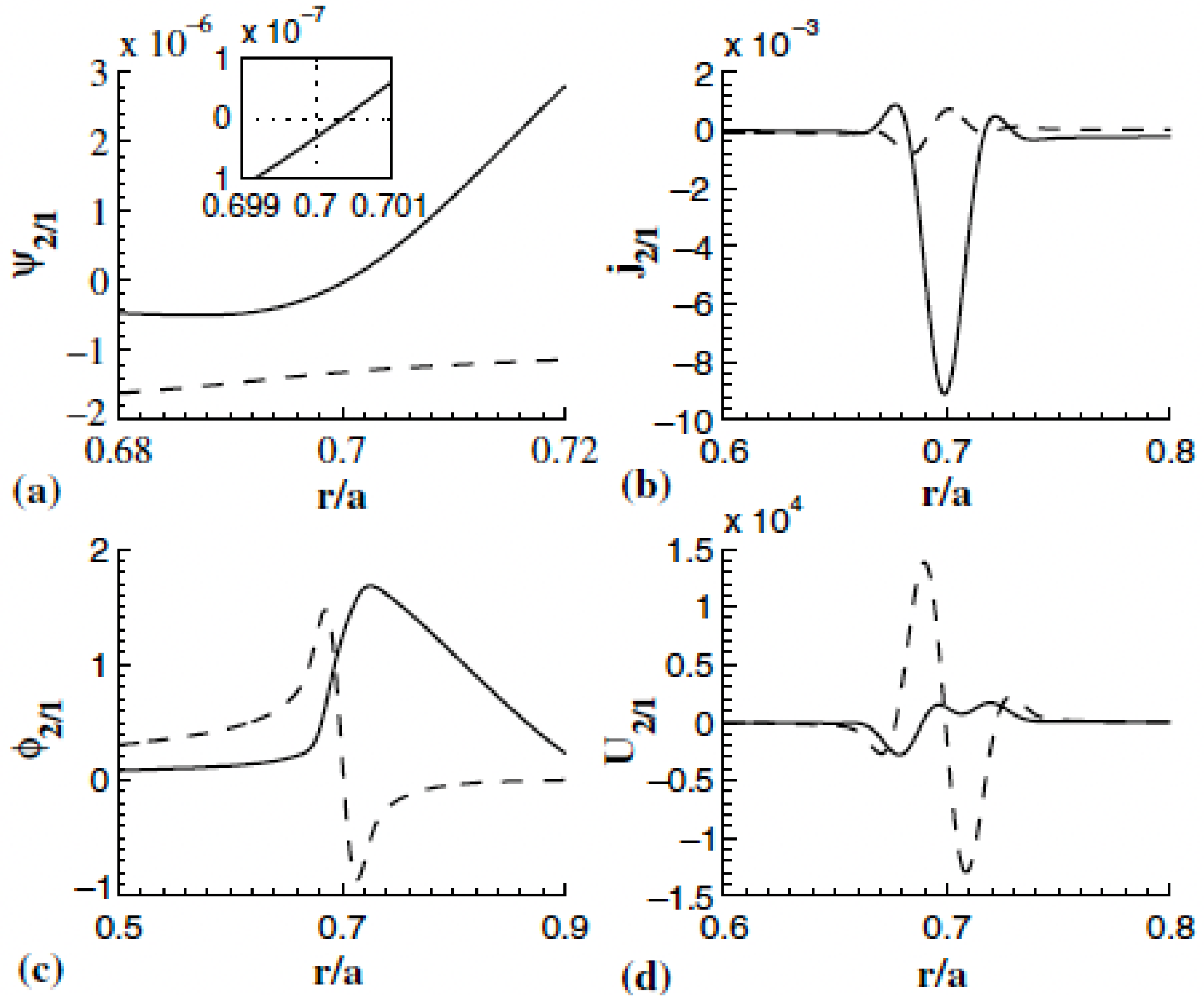

**Figure 8.** Corresponding to $\psi_a$=5×10$^{-5}aB_t$ in *small locked island regime* of figure 6(b), radial profiles of (a) $\psi_{2/1}$, (b) $j_{2/1}$, (c) $\phi_{2/1}$, and (d) $U_{2/1}$ in steady state. The solid (dashed) curve is the real (imaginary) part.

The *mode suppression regime* and *mode locking regime* have been analyzed before in reference 17 and 24. To authors' knowledge, the s*mall locked island regime has* not been found in previous theories. To have a look at the structure of the small locked island, corresponding to figure 6(b) with $\psi_a$=5×10$^{-5}$$aB_t$ in the *small locked island regime*, radial profiles of (a) the helical flux $\psi_{2/1}$, (b) current density $j_{2/1}$, (c) stream function $\phi_{2/1}$, and (d) plasma vorticity $U_{2/1}$ in steady state are presented in figure 8, where the subscript 2/1 indicates the $m/n$=2/1 component. The local profile of the real part of $\psi_{2/1}$ around the $q$=2 surface is shown at the top corner of figure 8(a). It is seen that the real part of $\psi_{2/1}$ is negative inside the $q$=2 surface at $r$=0.7$a$, showing the structure of a kink mode rather than a tearing mode. The imaginary part of $\psi_{2/1}$ is negative, and its amplitude is much larger than that of the real part, indicating that the small island is locked in the third quadrant, and the phase difference between the island and the applied RMP is slightly larger than π/2. With further increasing $\psi_a$ in the *small locked island regime*, the phase difference approaches π/2. This is different from the usual mode locking for which the phase difference is well below π/2 [17, 24]. The real part of $\phi_{2/1}$ is positive on both side of the $q$=2 surface. The current density perturbation $j_{2/1}$ in figure 8(b) and plasma vorticity $U_{2/1}$ in figure 8(d) are localized at the rational surface as expected.

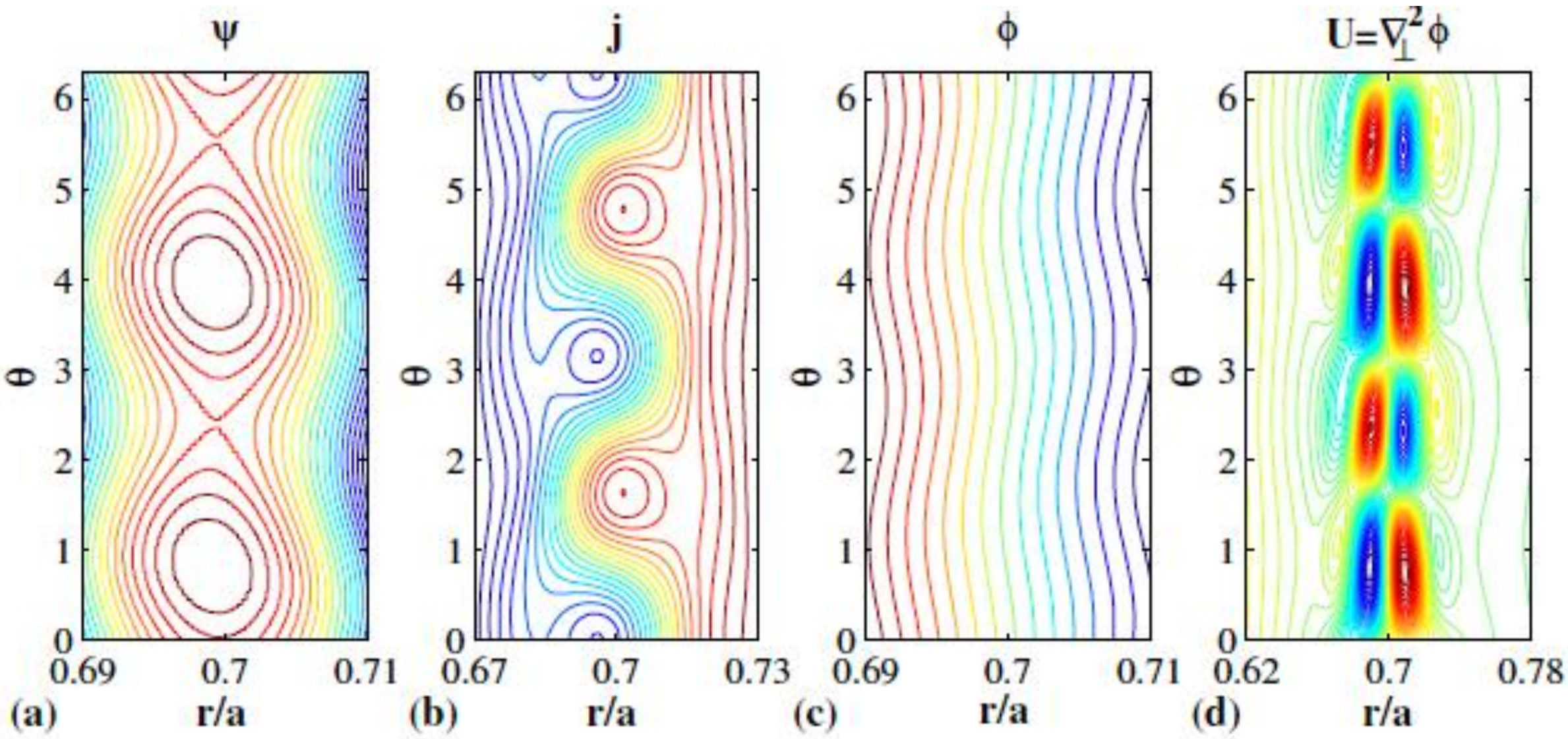


**Figure 9.** Corresponding to figure 8, contour plots for (a) helical magnetic flux $\psi$, (b) current density $j$, (c) stream function $\phi$, and (d) plasma vorticity $U$ in steady state.

Corresponding to figure 8, figure 9 shows the contour plots for (a) the helical magnetic flux $\psi$, (b) current density $j$, (c) stream function $\phi$, and (d) plasma vorticity $U$ in steady state. The small locked island is clearly seen from figure 9(a). From figure 9(b) it is seen that there is a radial shift between the maximum and the minimum of the current density perturbation, and the maximum and the minimum are not located at the o- or x-point of the island. Distortion from flow shear has been expected and predicted theoretically and observed experimentally in [19, 20]. Figures 9(c) and 9(d) indicate that the small locked island decouples from the plasma flow, and there is strong plasma vorticity in the island region.

For even larger original mode frequencies, the normalized mode frequency $\omega_p/\omega_0$ and island width $w/a$ at nonlinear saturation are shown as a function of $\psi_a$ in figure 10 for $\omega_0=1.6\times10^4/\tau_R$ (a) and $\omega_0=3.2\times10^4/\tau_R$ (b). For a larger $\omega_0$, the mode suppression regime and the small locked island regime become wider and extend to a larger value of $\psi_a$, indicating the increasing stabilizing role of the RMP for a larger plasma rotation frequency. One can see from figure 10(a) again that the mode frequency increases before entering into the small locked island regime, similar to that shown in figure 6(b).

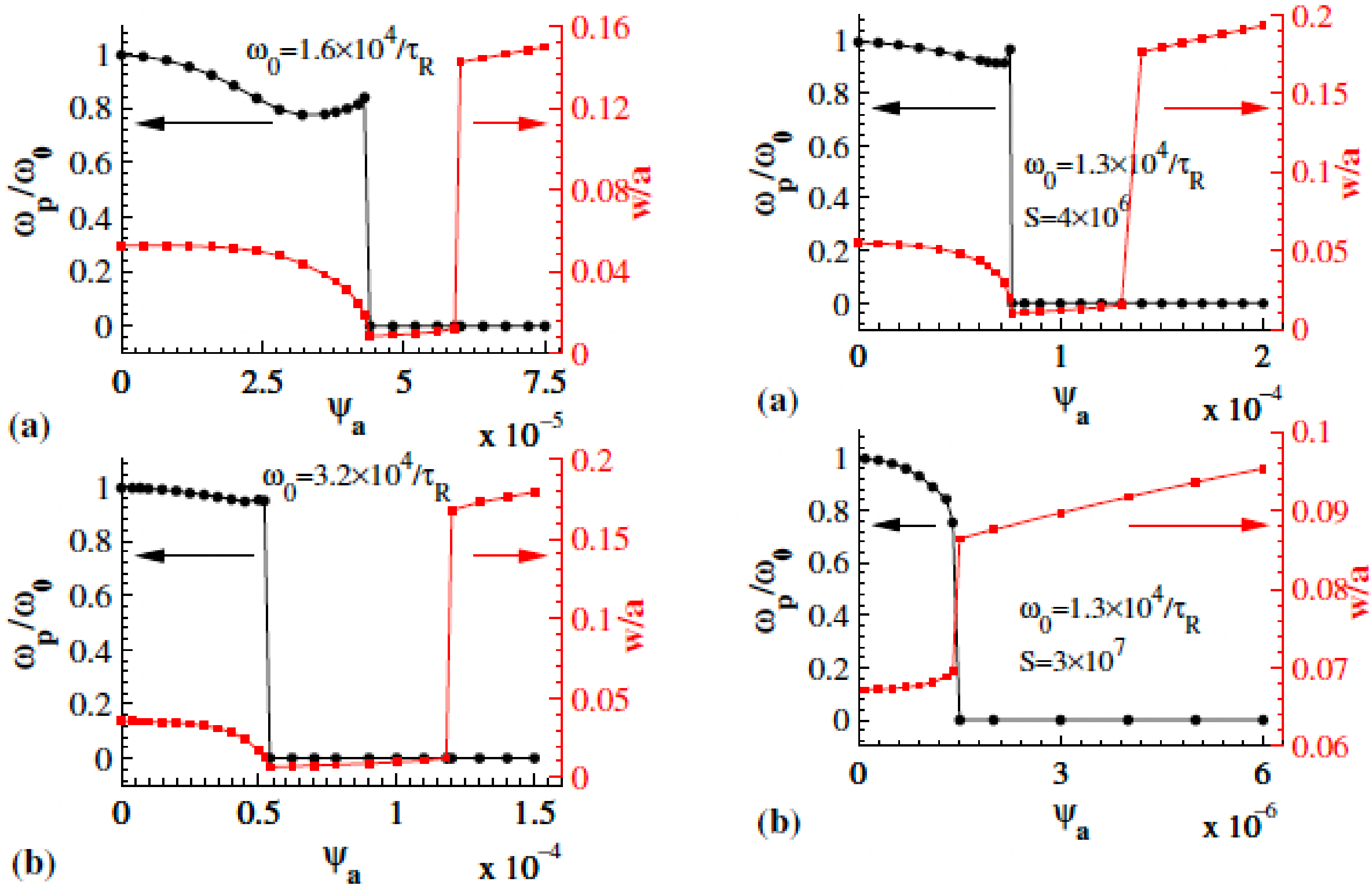


**Figure 10.** Normalized mode angular frequency $\omega_p/\omega_0$ and island width $w/a$ at nonlinear saturation versus $\psi_a$ for (a) $\omega_0=1.6\times10^4\tau_R$ and (b) $\omega_0=3.2\times10^4/\tau_R$.

**Figure 11.** Normalized mode angular frequency $\omega_p/\omega_0$ and island width $w/a$ at nonlinear saturation versus $\psi_a$ for (a) $S=4\times10^6$ and (b) $S=3\times10^7$, with $\omega_0=1.3\times10^4/\tau_R$.

To look into the effect of the magnetic Reynolds number S, in figure 11 the normalized mode frequency and island width $w/a$ at nonlinear saturation are shown as a function of $\psi_a$ for $\omega_0=1.3\times10^4/\tau_R$ with $S=4\times10^6$ (a) and $3\times10^7$ (b). For a smaller S value ($4\times10^6$), the mode stabilization by RMP is seen for $\psi_a<1.3\times10^{-4}aB_t$. For a larger S value ($3\times10^7$), however, no mode stabilization is observed. The mode frequency decreases with increasing $\psi_a$, while the island width slightly increases before entering into the mode locking regime, similar to those shown in figure 6(a). This suggests that the stabilizing effect comes from the plasma inertia or viscosity. It is seen from equation (8) that the plasma inertia and viscosity become less important for a larger S value. It should be mentioned that, as the resistive time is fixed in our calculations, a larger S value corresponds to a smaller Alfvén time here.

When further increasing the value of the magnetic Reynolds number to $S=1\times10^8$, the normalized mode frequency and island width at nonlinear saturation are shown as a function of $\psi_a$ for $\omega_0=3.2\times10^4/\tau_R$ in figure 12(a) and $6.3\times10^4/\tau_R$ in figure 12(b). For $\omega_0=3.2\times10^4/\tau_R$, no mode suppression by the RMP is found due to the large S value. Increasing the original mode frequency to $\omega_0=6.3\times10^4/\tau_R$, the mode suppression by the RMP appears again. It is seen that for a larger S value (smaller Alfvén time), the required plasma rotation frequency is higher in order to have the stabilizing effect by RMP.

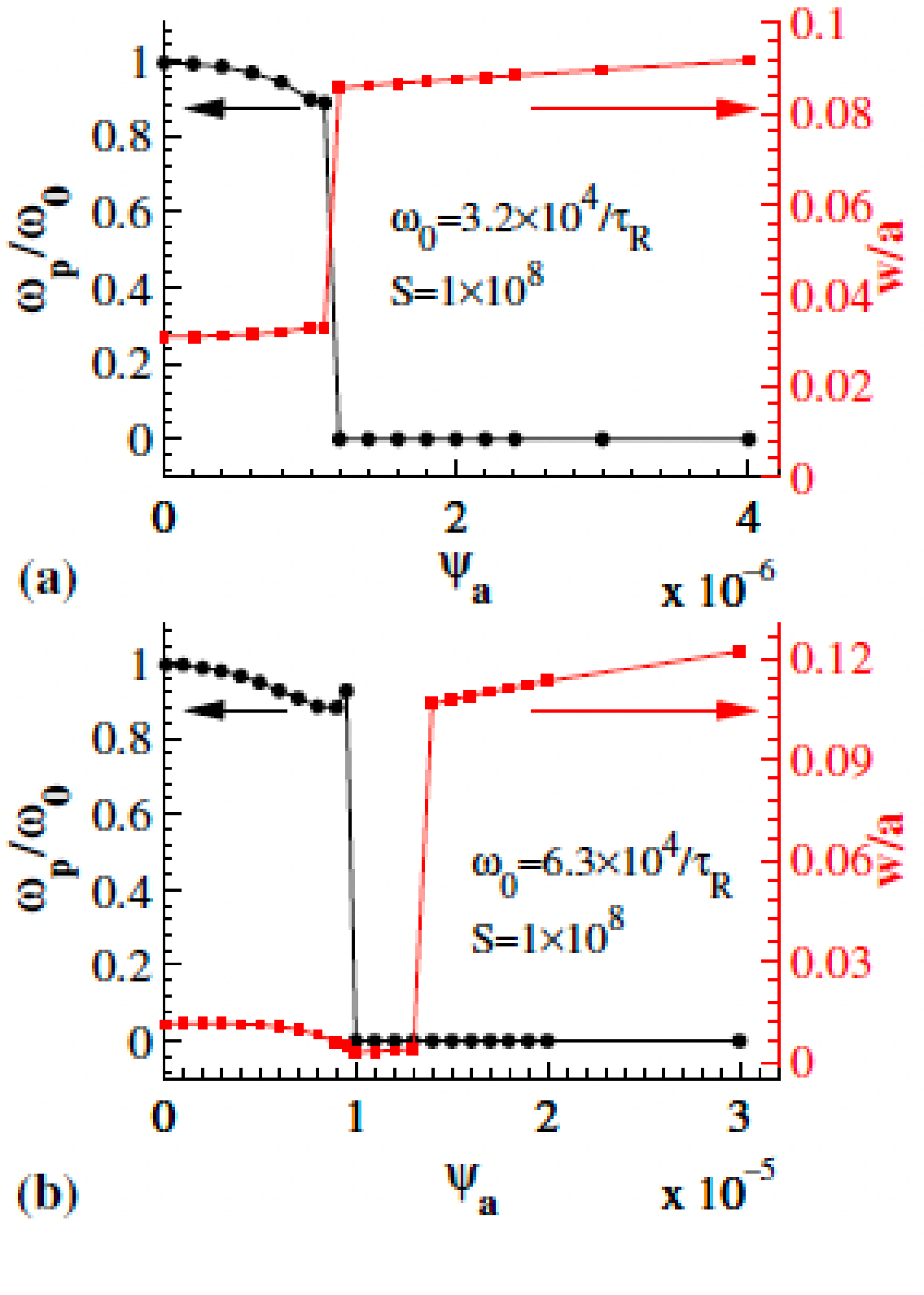


**Figure 12.** Normalized mode angular frequency $\omega_p/\omega_0$ and island width $w/a$ at nonlinear saturation versus $\psi_a$ for $S=1\times10^8$ with (a) $\omega_0=3.2\times10^4/\tau_R$ and (b) $\omega_0=6.3\times10^4/\tau_R$.

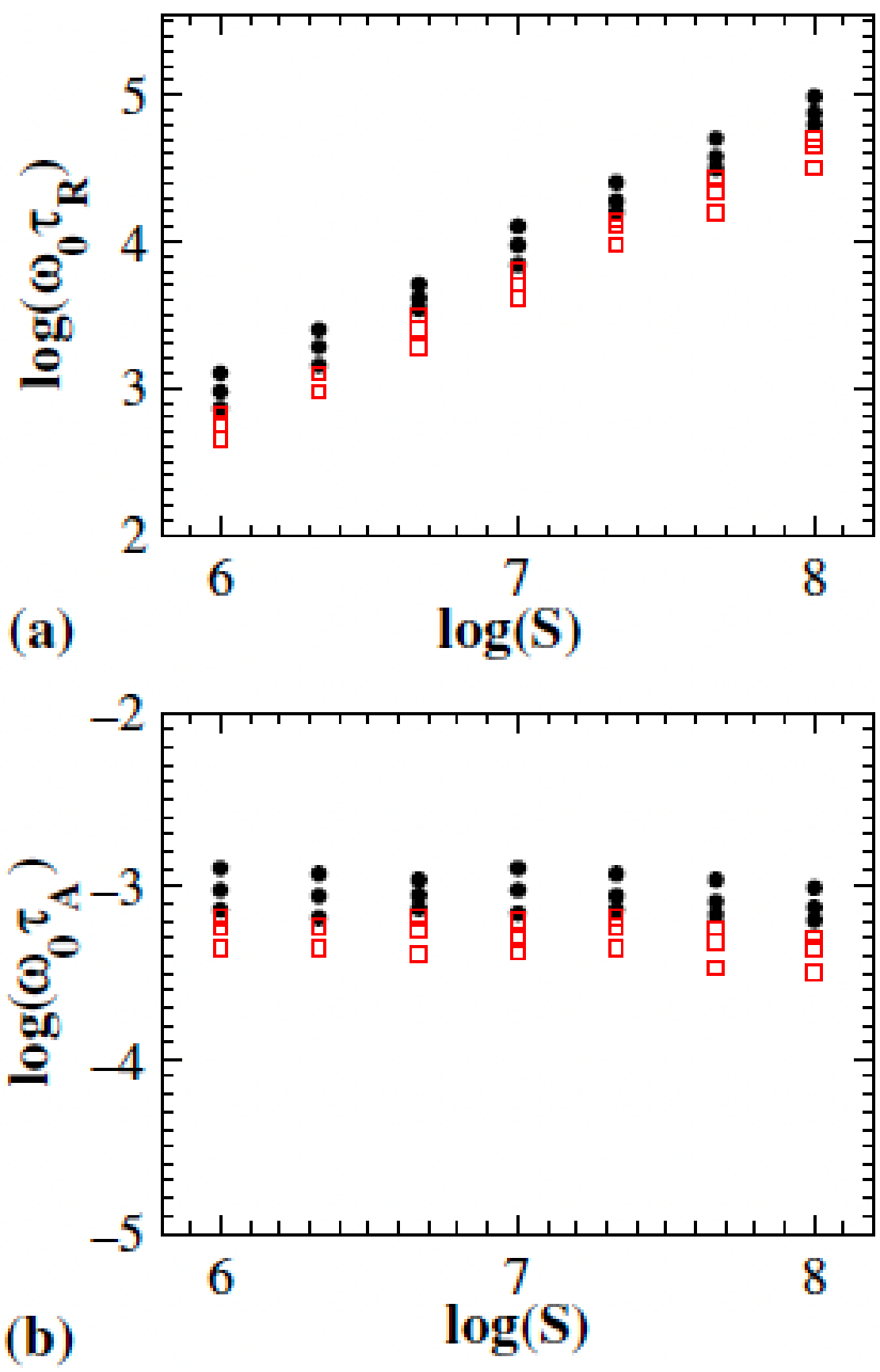


**Figure 13.** (a) Required normalized original mode angular frequency, $\omega_0\tau_R$, for mode stabilization by RMP versus magnetic Reynolds number S. The solid circles (empty squares) show the cases for which there is (no) mode stabilization. (b) Same as (a) except that the vertical axis is $\log(\omega_0\tau_A)$.

Figure 13(a) shows the required normalized original mode angular frequency $\omega_0\tau_R$ for mode stabilization by RMP versus magnetic Reynolds number S in the $\log(\omega_0\tau_R)$~log(S) plane. The solid circles (empty squares) show the case for which there is (no) mode stabilization. The required $\omega_0\tau_R$ for mode stabilization is proportional to S. When the mode angular frequency is normalized to the Alfvén time $\tau_A$ as shown in figure 13(b), however, it is seen that the required $\omega_0\tau_A$ for mode stabilization is nearly a constant for different S values. The critical value is

$\log(\omega_{0c}\tau_A)$=-3.18, and above which the 2/1 tearing mode can be stabilized by RMPs of moderate amplitude. Below this value no mode stabilization by RMPs is found. The value of $\log(\omega_{0c}\tau_A)$=-3.18 corresponds to $\omega_{0c}\tau_A=6.6\times10^{-4}$ or $f_{0c}=\omega_{0c}/2\pi=4.77$ kHz for typical J-TEXT experimental parameters ($\tau_A=2.2\times10^{-8}$s), being larger than the electron diamagnetic drift frequenc (~1.5kHz).

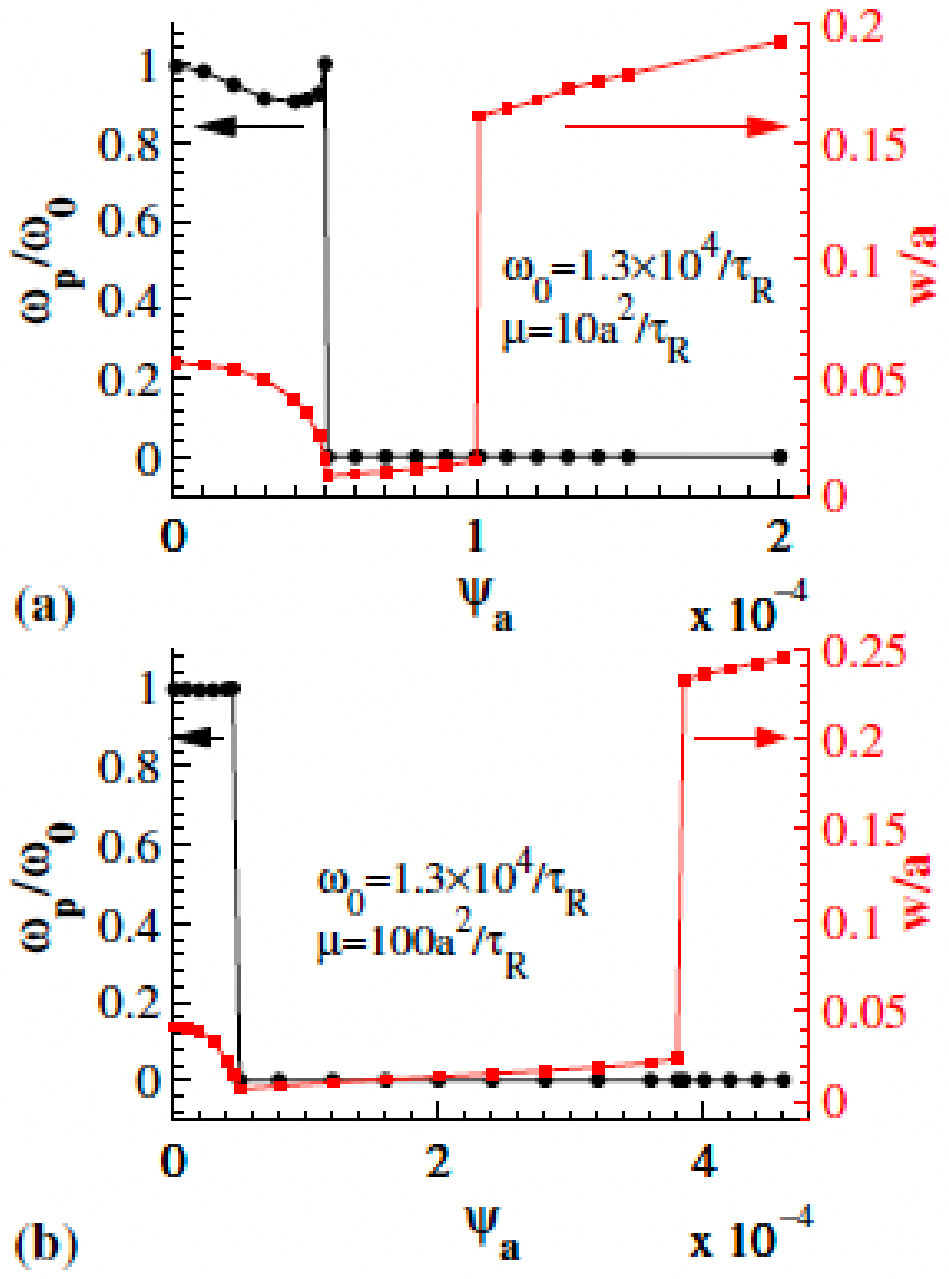


**Figure 14.** Normalized mode frequency $\omega_p/\omega_0$ and island width $w/a$ at nonlinear saturation versus $\psi_a$ for $\omega_0=1.3\times10^4/\tau_R$ with (a) $\mu=10a^2/\tau_R$ and (b) $\mu=100a^2/\tau_R$.

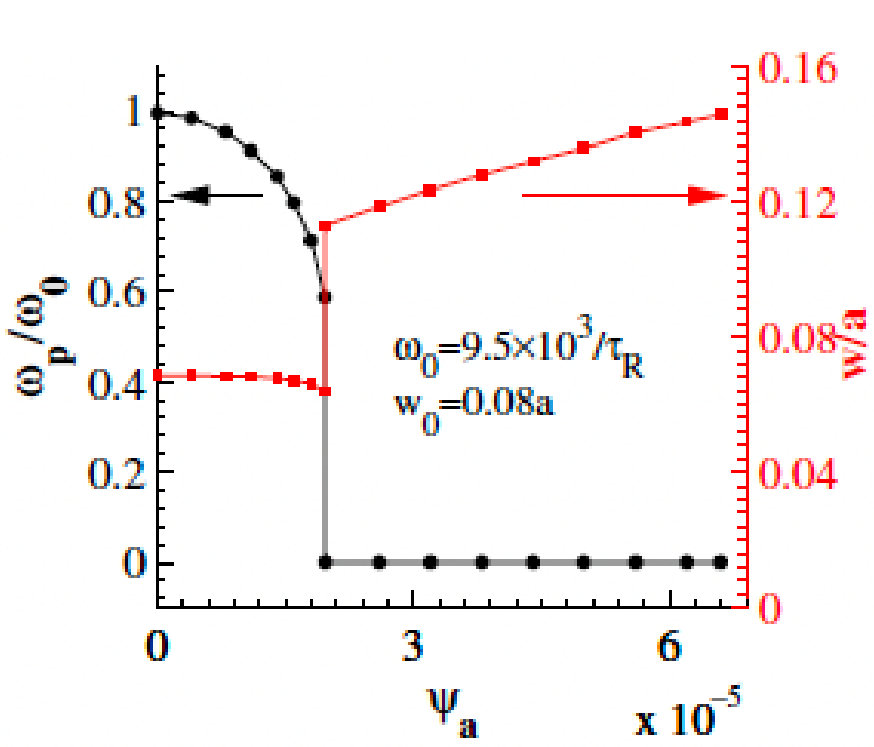


**Figure 15.** Normalized mode frequency, $\omega_p/\omega_0$, and island width, $w/a$, at nonlinear saturation versus $\psi_a$ with the initial island width $w_0=0.08a$. The other input parameters are the same as those for figure 6(b). In this case essentially no mode stabilization is observed as the initial mode amplitude is too large.

To study the effect of plasma viscosity, in figure 14 the normalized mode frequency and island width at nonlinear saturation are shown as a function of $\psi_a$ for $\omega_0=1.3\times10^4/\tau_R$ with two different values of the plasma viscosity, $\mu=10a^2/\tau_R$ (a) and $100a^2/\tau_R$ (b). For a larger value of $\mu$, the small island locking regime becomes wider, showing the increasing stabilizing effect by RMP for a larger plasma viscosity. If a too small plasma viscosity is used, the local plasma rotation velocity at the resonant surface is brought to zero at small RMP amplitude, resulting in mode locking before mode stabilization with increasing $\psi_a$.

The saturated island width also affects the result. This could be found from numerical calculations by using different initial island width at $t$=0 for the input, since the RMP is applied from $t$=0. For figures 6-14 $w_0=0.04a$ is taken. Figure 15 is for a larger initial island width, $w_0=0.08a$, while the other input parameters are the same as those for figure 6(b). In this case essentially no mode stabilization by RMP is found. Mode locking happens as $\psi_a$ increases to $1.97\times10^{-5}aB_t$. Detailed numerical calculations reveal that mode stabilization by RMP is possible

in this case only for $w_0$<0.057$a$. It is found from further calculations that the maximum initial island width to still has mode stabilization by RMP increases for a larger plasma rotation frequency and viscosity or for a lower S value. This could be understood since the mode locking threshold is proportional to the plasma viscosity and mode frequency but inversely proportional to the square of the Alfvén velocity and the island width [36]. The local plasma rotation velocity at the resonant surface is more easily brought to zero by RMP for a large initial island and S value and/or for a smaller plasma rotation frequency and viscosity.

## 4. Discussion and summary

It is well known that RMPs (or error field) can deteriorate tokamak plasma performance by causing mode penetration or mode locking [17, 23-33]. On the other hand, they can also have beneficial effects such as stabilizing ELMs [43-45]. An interesting experimental phenomenon has been observed on J-TEXT tokamak as well as on some other tokamaks is the stabilization of tearing mode by static RMPs of moderate amplitude [24, 39-42], providing an opportunity for comparison with theoretical results. Our numerical results, obtained by directly solving nonlinear reduced MHD equations, approximately agree with experimental observations on J-TEXT tokamak, showing that the mode stabilization by RMPs exists for a sufficiently high island rotation frequency and low Alfvén velocity. As strong plasma flow exists in the pedestal region of H-mode plasmas, our results suggest that RMPs of appropriate amplitude and mode numbers can stabilize local magnetic islands there. In the presence of a weak error filed (or RMPs), the tearing mode is more stable for a higher plasma rotation frequency, being in line with the experimental results [19-22]

Our numerical results can be partly explained by the analytical model presented in Ref. 24, in which the change in the mode angular frequency, $\Delta\omega$, by RMPs is taken into account, leading to a modified Rutherford equation (equation B.1 in Ref. 24),

$$r_s \frac{dw}{\eta dt} \propto r_s \Delta' + c \frac{\tau_A^2 F(\omega)}{(ns)^2 (w/r_s)^2}, \tag{10a}$$

where $\Delta'$ is the tearing mode stability index, $r_s$ is the minor radius of the resonant surface, $c$ is a negative numerical constant, $s=rq'/q$ calculated at $r=r_s$, and $F(\omega)=(\Delta\omega/2\pi)^2$. The second term on the right hand side of Eqn. (10a) describes the stabilizing effect by RMPs due to the changed mode frequency by RMPs [24]. Even if there are no applied RMPs, there is a similar ion polarization current contribution due to the plasma rotation and diamagnetic drift, which leads to a threshold for the onset of neoclassical tearing modes for sufficiently high $\beta$ plasmas [11-16].

Equation (10a) predicts that the stabilizing effect is larger for a smaller island width or larger $\tau_A$ value, being in agreement with our numerical results. When the island saturates (d$w$/d$t$=0), Eqn. (10a) leads to

$$\left(\frac{w_{sat}}{r_s}\right)^2 = -\left(\frac{\tau_A}{ns}\right)^2 \frac{cF(\omega)}{r_s\Delta'}.$$

(10b)

Eqn. (10b) shows that $(w_{sat}/r_s)2$ depends on the square of the product $\tau_A$ and $\Delta\omega$ divided by a current profile factor that determines tearing stability.

However, some features found from numerical results are not included in equation (10a), such as the stabilizing effect of the plasma viscosity [14]. One can see from figure 14 that for a large plasma viscosity, the mode frequency changes little with the decrease in the island width. Also, figure 11(b) and 12(a) indicate that the RMP is destabilizing for a high Alfvén velocity or low mode frequency. It is shown in Ref. [14] that the stabilizing effect by RMPs is due to the plasma viscosity rather than the changed plasma inertia. Further analytical work is required to have an improved formula and comparison with numerical results. The results shown in section 3.2 suggest that the key mechanism to hinder the mode stabilization by RMP is the mode locking, which corresponds to a zero plasma rotation velocity at the resonant surface. For a higher plasma rotation frequency and/or a lower Alfvén velocity, the mode locking threshold is higher, and therefore mode stabilization is possible with increasing RMP amplitude before reaching the mode locking threshold [36].

It has been shown that when the plasma rotation frequency is sufficiently high, applied static RMPs can lead to Alfvén resonances, which shield the forced magnetic reconnection due to RMPs [51, 52]. In the linear phase two different regimes have been identified, the magnetic reconnection regime existing for a low plasma rotation frequency and the Alfvén resonance regime for a high frequency [51, 52]. In our numerical results the 2/1 tearing mode is unstable even without RMPs and saturates at large amplitude in the nonlinear phase. In this case the mode behaviour is dominated by magnetic reconnection. However, it can not be ruled out that the Alfvén resonance has an effect on the mode stabilization. Analytical formula for the role of the Alfvén resonances in the nonlinear phase is required for comparison with numerical results.

When the plasma rotation frequency is not high enough, the small locked island regime only exists for a narrow range of the RMP amplitude, as can be seen from figure 6(b). This makes it difficult to identify it experimentally. As shown in figure 3, the 2/1 mode enters into the mode locking regime 6ms later after it is stabilized (possibly in the small locked island regime). Future experiments with more careful adjustment of the waveform and amplitude of the saddle coil current are required to see if the small locked island regime can be observed. It was reported that small locked island was found on T-10 tokamak experiments [53].

In summary, the *m*/*n*=2/1 tearing modes are observed to be (partly) stabilized by externally applied static RMPs of moderate amplitude in J-TEXT tokamak experiments. With realistic plasma parameters as input, the numerical results obtained from nonlinear reduced MHD equations approximately agree with experimental observations. In addition, it is found from numerical calculations that the mode stabilization by RMP is possible for a sufficiently high island rotation frequency and a low Alfvén velocity. A larger plasma viscosity enhances the mode stabilization.

## Acknowle[Qiming, 2012 #307]ments

One of the authors (Q. Hu.) is grateful to Mr. Wei Jin and Ms. Yan He for useful discussions and to Dr. S. Elgriw and C. Xiao of the University of Saskatchewan for their help in experimental data analysis. This work is supported by the Ministry of Science and Technology (Contract No. 2010GB107004, 2011GB109001, 2008CB717805, and 2009GB105003) and the Chang-Jiang scholar project of the Ministry of Education, China.